\documentclass[12pt]{article}
\usepackage{amsmath,amssymb,amsthm,amscd,amsfonts,geometry,hyperref,indentfirst,color,setspace}

\numberwithin{equation}{section}

\def\be{\begin{equation}}
\def\ee{\end{equation}}
\def\ba{\begin{array}}
\def\ea{\end{array}}
\def\bn{\begin{aligned}}
\def\en{\end{aligned}}
\def\bsub{\begin{subequations}}
\def\esub{\end{subequations}}

\def\p{{\partial}}
\def\pb{{\bar\partial}}

\def\a{{\alpha}}
\def\b{{\beta}}

\def\g{{\gamma}}
\def\d{{\delta}}
\def\e{{\varepsilon}}

\def\z{{\zeta}}
\def\h{{\eta}}
\def\th{{\theta}}

\def\l{{\lambda}}
\def\lb{{\bar\lambda}}

\def\P{{\Pi}}

\def\s{{\sigma}}
\def\t{{\tau}}

\def\F{{\Phi}}

\def\w{{\omega}}
\def\W{{\Omega}}

\begin{document}

\begin{titlepage}

\title{
\vskip-40pt
\begin{flushright}
{\small ICTP-SAIFR/2013-012}\\
~\\
~\\
\end{flushright}
{\bf Pure Spinor $b$-ghost in a Super-Maxwell Background}
~\\
~\\
\author{Ilya~Bakhmatov\footnote{\tt ilya@ift.unesp.br}~~and~Nathan~Berkovits\footnote{\tt nberkovi@ift.unesp.br}
~\\
~\\
{\it ICTP South American Institute for Fundamental Research}\\
{\it Instituto de F\'{\i}sica Te\'orica, UNESP - Univ. Estadual Paulista }\\
{\it Rua Dr. Bento T. Ferraz 271, 01140-070, S\~ao Paulo, SP, Brasil}
~\\
\date{}
}}

\maketitle

\begin{abstract}
\noindent
In the pure spinor formalism for the superstring, the $b$-ghost is a composite operator satisfying $\{Q,b\}=T$ where $Q$ is the pure spinor BRST operator and $T$ is the holomorphic stress tensor. The $b$-ghost is holomorphic in a flat target-space background, but it is not holomorphic in a generic curved target-space background and instead satisfies $\pb b = [Q, \Omega]$ for some $\Omega$. In this paper, $\Omega$ is explicitly constructed for the case of an open superstring in a super-Maxwell background.
\end{abstract}
~\\
~\\

\thispagestyle{empty}
\end{titlepage}

\tableofcontents

\section{Introduction}

The description of quantum superstrings using the pure spinor formalism has several advantages over the Ramond-Neveu-Schwarz formalism. Since spacetime supersymmetry is manifest, there is no sum over spin structures and multiloop amplitudes (such as the recent three-loop computation of \cite{Gomez:2013sla}) are easier to compute. Furthermore, fermionic and bosonic states can be treated symmetrically, so Ramond-Ramond backgrounds such as $AdS_5\times S^5$ can be described \cite{Berkovits:2004xu}. Although there are many similarities of
the pure spinor formalism with the Green-Schwarz formalism, the pure spinor formalism in a flat target-space background has a quadratic worldsheet action, so it has the advantage over the Green-Schwarz formalism that covariant quantization is straightforward.

An unusual feature of this approach is that is does not start with a reparameterization invariant worldsheet formulation. Instead, the formalism starts with a worldsheet action in conformal gauge and the BRST symmetry is postulated rather than derived from fixing a gauge symmetry. Correspondingly, there are no natural Faddeev-Popov $(b,c)$-ghosts whose zero
modes normally appear in the integration measure on the moduli space of the Riemann surface when computing string scattering amplitudes. Nevertheless, one can define the $b$-ghost in the pure spinor formalism as a composite operator satisfying $\{Q,b\}=T$, where $Q$ is the BRST operator and $T$ is the holomorphic energy-momentum tensor (we will ignore the antiholomorphic sector for simplicity). 

In a flat target-space background, this composite $b$-ghost was explicitly constructed in \cite{Berkovits:2005bt} using a chain of holomorphic spacetime supersymmetric operators \cite{Berkovits:2006vi}\cite{Oda:2007ak}. Although the composite operator is complicated and nilpotency has only recently been verified \cite{Chandia:2010ix}\cite{Jusinskas:2013yca}, the construction is simplified by introducing a twisted RNS-like variable \cite{Berkovits:2013pla} which may eventually help in understanding its structure.

Because the $b$-ghost is necessary for computing loop scattering amplitudes, it is important to construct the $b$-ghost in curved target-space backgrounds. Unlike in bosonic or RNS string theory, the integrated vertex operators used to deform the target-space background in the pure spinor formalism are not in Siegel gauge, i.e. they have singular OPE's with the $b$-ghost. This implies that after deforming the background, the composite operator for the $b$-ghost will no longer be holomorphic but instead will satisfy 
\be\label{inone}
\pb b = \left[ Q, \W\right]
\ee
for some
operator $\W$ defined up to the equivalence relation $\W \sim \W + \{Q,\Lambda\}$. In \cite{Berkovits:2010zz}, the operator $\W$ was constructed for the case of the Type IIB superstring in an $AdS_5 \times S^5$ background, and in this paper, the operator
$\W$ will be constructed for the open superstring in a super-Maxwell background. It is expected that the results of this paper for the open superstring background will be useful for
understanding the structure of the $b$-ghost in a general Type II closed superstring background.

After reviewing in section \ref{backgr} the pure spinor description of an open superstring in a supersymmetric Maxwell background, we construct in section \ref{const} an operator $\W$ satisfying (\ref{inone}). In terms of the super-Maxwell vertex operator $V$,
 \be\label{intwo}
 \W = b_{-1} b_0 V -{1\over 2} \p (b_{-1} b_1 V)
 \ee
where $b_n {\cal O}$ is the residue of the pole of order $(n+2)$ in the OPE of the $b$-ghost with ${\cal O}$, i.e. 
\be
b_n {\cal O} (z) = \frac{1}{2\pi i} \oint dy\, (y-z)^{n+1}\, b(y)\, {\cal O} (z).
\ee
Since the $b$-ghost
contains poles up to order $(\bar\lambda\lambda)^{-4}$ in the pure spinor ghosts, the expression of (\ref{intwo}) for $\W$ will generically contain poles up to order $(\bar\lambda\lambda)^{-8}$. However, it will be shown by explicit construction that a representative in the cohomology of $\W$ can be chosen to have poles only up to order $(\bar\lambda\lambda)^{-4}$ in the pure spinor ghosts. The explicit expression for $\W$ in terms of the supersymmetric operators appearing in the $b$-ghost is given in (\ref{final}), and for the special case where the background electromagnetic field strength $F_{mn}$ is constant, $\W$ simplifies to $\W = \frac14 F_{mn} \left(\bar\lambda \g^{mn}{{\p}\over{\p r}} \right) b $. 

\section{Review of super-Maxwell background}
\label{backgr}

To describe the open superstring in a supersymmetric Maxwell background, one adds to the flat space action a massless vertex operator integrated over the boundary of the open string worldsheet \cite{Berkovits:2002ag}\cite{Berkovits:2002zk}. Taking for simplicity that the worldsheet has only one boundary at $\s = 0$, the action in the pure spinor formalism is
\be\label{action}\bn
S &= \int d\s d\t \left( \frac12 \p x^m \pb x_m + p_\a \pb \th^\a + \w_\a \pb \l^\a +\bar w^\a \pb \bar\l_\a  + s^\a \pb r_\a \right) \\&+ \int d\s d\t \,\d(\s) \left( \p\th^\a A_\a + \P^m A_m + d_\a W^\a + \frac12 N_{mn} F^{mn} \right).
\en\ee
The action is written in terms of the following worldsheet variables: $x^m$ and $\th^\a$ are the standard $D=10$ superspace variables; $p_\a$ is the conjugate variable for $\th^\a$; $\l^\a$ is a bosonic pure spinor variable subject to the constraint $\l\g^m\l =0$ with its conjugate $\w_\a$; $\bar\l_\a$ is a bosonic pure spinor variable subject to the constraint $\bar\l\g^m\bar\l =0$ with its conjugate $\bar\w^\a$;
and $r_\a$ is a fermionic variable subject to the constraint $\bar\l\g^m r=0$ with its conjugate $s^\a$. Note that $(\bar\l_\a, \bar\w^\a, r_\a, s^\a)$ are non-minimal variables which do not appear in physical vertex operators but which are necessary for constructing the $b$-ghost. Although one has the freedom to add BRST-trivial terms involving non-minimal variables to the vertex operator, we will not do so in this paper.

Throughout this paper, we will ignore the antiholomorphic variables by using
the 
usual ``doubling trick'' where the holomorphic and antiholomorphic variables of the
open string on the interval $0\leq\sigma\leq \pi$ are replaced by holomorphic variables
of a periodic closed string with $0\leq\sigma<2\pi$. In this description, the antiholomorphic
variables of the open string on the interval $0\leq\sigma\leq \pi$ are described as holomorphic
variables of the closed string at the position $2\pi - \sigma$, and $D9$-brane boundary
conditions imply that the closed string variables are continous at $\sigma=0$ and $\sigma =\pi$. For example, $\theta^\a_{closed}(\s) = \theta^\a_{open}(\s)$ when $0\leq\s\leq\pi$ and
$\theta^\a_{closed}(\s) = \bar\theta^\a_{open}(2\pi -\s)$ when $\pi\leq\s\leq2\pi$.

The vertex operator in (\ref{action}) involves the background fields of the $D=10$ supersymmetric Maxwell theory: $A_\a$ and $A_m$ are the spinor and vector gauge superfields, and $W^\a$ and $F_{mn} = \p_m A_n - \p_n A_m$ are the superfield strengths. They are constrained by the equations \cite{Siegel:1978yi}\cite{Witten:1985nt}:
\be\label{max}\bn
\g^m_{\a\b} A_m &= D_\a A_\b + D_\b A_\a, \\
\g_{m\a\b} W^\b &= D_\a A_m - \p_m A_\a, \\
D_\a W^\b &= \frac14 {(\g_{mn})_\a}^\b F^{mn}.
\en\ee
which imply the familiar super-Maxwell equations of motion for the field strengths:
\be\label{max}
\p_m F^{mn} = 0, \qquad \g^m_{\a\b} \p_m W^\b = 0, \qquad D_\a F_{mn} = \p_m (\g_n W)_\a - \p_n (\g_m W)_\a.
\ee
We have introduced the supersymmetric derivative $D_\a = \p_\a + \frac12 (\g^m \th)_\a \p_m$ as well as the bosonic and fermionic supersymmetric momenta,  $\P^m = \p x^m + \frac12 \th\g^m\p\th$ and $d_\a = p_\a - \frac12 \left( \p x^m + \frac14 \th\g^m\p\th \right) (\g_m\th)_\a$.
Because of the constraint $\l\g^m\l = 0$, the pure spinor action possesses a gauge symmetry transformation 
\be\label{gaugew}
\d\w_\a = \Lambda^m (\l\g_m)_\a
\ee
for any $\Lambda^m$. Gauge invariant pure spinor Lorentz and ghost number currents are given by
\be
N_{mn} = \frac12 \l^\a {(\g_{mn})_\a}^\b \w_\b, \qquad J = -\l^\a \w_\a,
\ee
and the left-moving energy-momentum tensor is the same as in a flat background:
\be\label{T}
T = -\frac12 \P^m \P_m - d_\a \p\th^\a - \w_\a \p\l^\a - \bar w^\a \p \bar\l_\a - s^\a \p r_\a.
\ee

The vertex operator 
\be\label{V}
V= \p\th^\a A_\a + \P^m A_m + d_\a W^\a + \frac12 N_{mn} F^{mn}
\ee
in (\ref{action}) satisfies $\left[ Q, \int dz V \right] =0$ under the action of the standard flat space BRST operator $Q = {1\over{2\pi i}}\oint (\l^\a d_\a -\bar w^\a r_\a)$ \cite{Berkovits:2005bt}. Thus the entire action is invariant under the flat space BRST transformation, and the usual construction of the $b$-ghost using the non-minimal pure spinor formalism \cite{Berkovits:2005bt} is still valid and the $b$-ghost satisfying $\{Q,b\}=T$ is:
\be\label{b}
b = s^\a \p\bar\l_\a + \frac{\bar\l_\a G^\a}{(\bar\l\l)} + \frac{\bar\l_\a r_\b H^{[\a\b]}}{(\bar\l\l)^2} - \frac{\bar\l_\a r_\b r_\g K^{[\a\b\g]}}{(\bar\l\l)^3} - \frac{\bar\l_\a r_\b r_\g r_\d L^{[\a\b\g\d]}}{(\bar\l\l)^4},
\ee
where\footnote{The antisymmetrization convention used in this article is without weight factors, e.g. $x^{[m}y^{n]} = x^m y^n - x^n y^m$.}
\be\label{chain}\bn
G^\a &= \frac12 \P^m (\g_m d)^\a - \frac14 N_{mn} (\g^{mn} \p\th)^\a - \frac14 J \p\th^\a,\\
H^{[\a\b]} &= \frac{1}{192} (\g^{mnk})^{\a\b} (d\g_{mnk} d + 24 N_{mn} \P_k),\\
K^{[\a\b\g]} &= -\frac{1}{96} (\g_m d)^{\left[\a\right.} (\g^{mnk})^{\left.\b\g\right]} N_{nk},\\
L^{[\a\b\g\d]} &= -\frac{1}{128} \frac{1}{4!} (\g_{mnp})^{\left[\a\b\right.} (\g^{pqr})^{\left.\g\d\right]} N^{mn} N_{qr}.\\
\en\ee

Although the operators  $H^{[\a\b]}, K^{[\a\b\g]}, L^{[\a\b\g\d]}$ of (\ref{chain}) are antisymmetric in
their spinor indices, it will be useful to define non-antisymmetric versions of these operators,
\be\label{nons}\bn
H^{\a\b} &= \frac{1}{16} N^{mn} \P^k (\g_m \h_{nk} + \g_{mnk})^{\a\b} - \frac{1}{32} \g_m^{\a\b} J \P^m -\frac{1}{16} (\g_m d)^\a (\g^m d)^\b,\\
K^{\a\b\g} &= {1\over{128}} \g_m^{\b\g} (\g^m d)^\a J + a_{mn}^{\a\b\g\d} N^{mn} d_\d,\\
L^{\a\b\g\d} &= c^{\a\b\g\d} J J + c_{mn}^{\a\b\g\d} N^{mn} J + c_{mnpq}^{\a\b\g\d} N^{mn}N^{pq},\\
\en\ee
where the coefficients ($a_{mn}^{\a\b\g\d}$, $c^{\a\b\g\d}$, $c_{mn}^{\a\b\g\d}$, $c^{\a\b\g\d}_{mnpq}$) can be computed using the Y-formalism \cite{Oda:2007ak}. 
The operators $G^\a, H^{\a\b}, K^{\a\b\g}, L^{\a\b\g\d}$
were first introduced in the ``minimal'' version of the $b$-ghost and satisfy the following BRST transformations \cite{Berkovits:2004px}:
\be\label{cep}\bn
\left\{ Q, G^\a \right\} &= \l^\a T_\mathrm{min}, \\
\left[ Q, H^{\a\b} \right] &= \l^\a G^\b + g^{((\a\b))}, \\
\left\{ Q, K^{\a\b\g} \right\} &= \l^\a H^{\b\g} + h_1^{((\a\b))\g} + h_2^{\a((\b\g))},\\
\left[ Q, L^{\a\b\g\d}  \right] &= \l^\a K^{\b\g\d} + k_1^{((\a\b))\g\d} + k_2^{\a((\b\g))\d} + k_3^{\a\b((\g\d))},\\
0 &= \l^\a L^{\b\g\d\e} +l_1^{((\a\b))\g\d\e} +l_2^{\a((\b\g))\d\e} +l_3^{\a\b((\g\d))\e} +l_4^{\a\b\g((\d\e))},
\en\ee
where $T_\mathrm{min}$ is given by the first three terms in (\ref{T}), and $g, h_i, k_j, l_k$ are operators which are symmetric and gamma matrix traceless in the pairs of indices indicated,
\be
f^{((\a\b))} = f^{((\b\a))}, \quad \g^m_{\a\b} f^{((\a\b))} =0.
\ee
All of these operators will cancel in the computations, so their explicit expressions will never be needed. Note that the last relation in (\ref{cep}) implies $\l^{\left[\a\right.} L^{\left. \b\g\d\e\right]}=0$.

\subsection{Equations of motion}

One approach to computing the effect of adding interaction terms such as the second line of (\ref{action}) to the action is to change the boundary conditions from Neumann $x^{m\prime} = 0$ to mixed and to leave unchanged the equations of motion. For example, for the bosonic string with action
\be
S=\int_\Sigma d\t d\s \,\frac12 \left( \dot x^m \dot x_m - x^{m\prime} x_m'\right) + \int_{\p\Sigma} d\t\, \dot x^m A_m,
\ee
variation with respect to $x^m$ implies (assuming just one boundary at $\s=0$ for simplicity)
\be\label{beom}
\ddot x^m - x^{m\prime\prime} + \d(\s) ( x^{m\prime} + \dot x_k F^{km}) = 0.
\ee
Using this approach, one keeps the equation of motion $\ddot x^m - x^{m\prime\prime} = 0$ and applies mixed boundary conditions $x^{m\prime} = F^{mn} \dot x_n$ at $\s=0$. In the pure spinor formalism for the open superstring, if one imposes analogous mixed boundary conditions on the worldsheet variables and requires that the holomorphic and antiholomorphic BRST currents coincide on the boundary, one finds that the background superfields in the vertex operator of (\ref{V}) must satisfy the
supersymmetric Born-Infeld equations  \cite{Berkovits:2002ag}. Using this approach, one would find that the holomorphic $b$-ghost of (\ref{b}) and the antiholomorphic $\bar b$ ghost do not coincide on the boundary, but satisfy
$b = \bar b + [Q,\W]$ for some $\W$.

A second approach to computing the effect of adding interaction terms is to leave unchanged the Neumann boundary conditions but modify the equations of motion by a term
proportional to $\d(\s)$. So for (\ref{beom}), the modified equation of motion is $\ddot x^m - x^{m\prime\prime}  = - \d(\s)  \dot x_k F^{km}$ where the interaction term has the effect of a force acting on the endpoint of the string similar to the Lorentz force term $F^{mn} \dot x_n$ of a point particle in an external field. In this approach (which we will follow here), 
holomorphicity of the BRST current implies the onshell equations for the background superfields and the $b$-ghost of (\ref{b}) in this background satisfies $\pb b =  [Q,\W]   \d(\s)$ for some $\W$. Throughout this paper, we will only compute to linearized order in the background superfields. So we will use the super-Maxwell equations as the onshell equations of the background instead of the full supersymmetric Born-Infeld equations.

Following this second approach and using a shorthand notation for the field strength with spinorial indices,
\be\label{rel}
{F_\a}^\b = \frac14 {(\g_{mn})_\a}^\b F^{mn} = D_\a W^\b,
\ee
the equations of motion for the worldsheet fields that follow from the action (\ref{action}) are:
\begin{gather}
\label{eom1}
\pb \l^\a = {F^\a}_\b \l^\b \d(\s), \qquad \pb \w_\a = {F_\a}^\b \w_\b\, \d(\s), \qquad \pb \th^\a = -W^\a \d(\s),\\
\bn
\pb\p x_m = &\Bigl[ F_{mn}\P^n + d_\a \p_m W^\a + \frac12 N_{pq} \p_m F^{pq} - \frac12 (\p\th\g_m W) \\ &+ \frac12 (\th\g_m \p W) \Bigr]\d(\s),
\en
\end{gather}
\be\bn
\pb p_\a = \biggl[ 
   					& -\frac12 \P^m (\g^n\th)_\a F_{mn} - \P^m (\g_m W)_\a + \frac12 \p x^m (\g_m W)_\a + d_\b \p_\a W^\b \\&- \frac12 N_{pq} \p_\a F^{pq}   + \frac38 (\g^m \p\th)_\a (\th\g_m W) - \frac48 (\g^m \th)_\a (\p\th\g_m W)   \\
					& + \frac18(\g^m \th)_\a (\th\g_m \p W) \biggr]\d(\s)
\en\ee

Using these formulae we can compute $\pb \P^m$ and $\pb d_\a$:
\begin{gather}
\pb \P_m = \d(\s) \left[ F_{mn}\P^n  + d_\a \p_m W^\a + \frac12 N_{pq} \p_m F^{pq} \right],\\
\label{eom2}
\pb d_\a = \d(\s) \left[ {F_\a}^\b d_\b - \frac12 N_{mn} D_\a F^{mn} \right].
\end{gather}
In the rest of the paper we will suppress the factor of $\d(\s)$ to keep the expressions simpler.

The simple form of $\pb\l^\a$ and $\pb\w_\a$ implies that $\pb J = 0$ and it is easy to calculate $\pb N_{mn} = F_{mk} {N^k}_n - F_{nk} {N^k}_m$. The $\p_m W^\a$ in the formulae above can also be expressed in terms of $D_\a F^{mn}$ using the relation that follows from (\ref{rel})
\be\label{dW=DF}
\p_m W^\a = \frac18 {\g_m}^{\b\g} D_\b {F_\g}^\a.
\ee

Using (\ref{rel}) and the property $D_\a D_\b + D_\b D_\a = \g^m_{\a\b} \p_m$ one can check that the BRST current $ \l^\a d_\a$ is holomorphic:
\be\bn
\pb (\l^\a d_\a) &= -\frac14 \l^\a (\l \g_{mn} \w) D_\a F^{mn} = -\l^\a \l^\b \w_\g D_\a {F_\b}^\g \\
				 &= -\w_\g \l^\a \l^\b D_\a D_\b W^\g = -\frac12 \w_\g (\l \g^m \l) \p_m W^\g = 0.
\en\ee
Holomorphicity of the worldsheet energy-momentum tensor (\ref{T}) is also easy to verify. Thus the relation $\{Q,b\} = T$ implies
\be
\{Q,\pb b\} = 0.
\ee
We would like to show that $\pb b = [Q,\W]$ for some $\W$. We will first consider the simple case of constant Maxwell field strength, and will then consider the more general case.

\section{Construction of $\Omega$}
\label{const}

Since $b(z)$ of (\ref{b}) is holomorphic in a flat background, $\pb b$ comes from poles in the OPE of $b(z)$ with the super-Maxwell vertex operator $V(y)$ of (\ref{V}). These poles are computed
by the commutator 
\be\label{comm}
\pb b = \left[{1\over{2\pi i}} \oint dy\, V(y),\, b(z) \right]
\ee
where $\oint dy$ is a contour integral of $y$ around $z$. Note that (\ref{comm}) is invariant under gauge transformations of the background superfields which transform $V(y) \to V(y) +\p_y \Lambda(y)$. So we need to construct an $\W$ such that
$ \left[{1\over{2\pi i}} \oint dy\, V(y) ,\, b(z) \right] = \left[Q,\W\right]$.
We shall begin by constructing $\W$ for the case of a constant background, and will then construct $\W$ for a general on-shell super-Maxwell background.

\subsection{$\Omega$ in constant electromagnetic field-strength}

When the electromagnetic field-strength of the background is constant, i.e. when 
$\p_m W^\a=0$ and $\p_m F_{nk} = 0$, the integrated vertex operator $V$ simplifies to $V = q_\a W^\a|_{\theta=0} - \frac12  M^{mn} F_{mn}$ where  
\be\label{qs}
q_\a = p_\a +  \frac12 \left( \p x^m + {1\over 12}\, \th\g^m\p\th \right) (\g_m\th)_\a
\ee
is the spacetime supersymmetry current and 
\be\label{Mconst}
M^{mn} = -\frac12 x^{[m} \p x^{n]} + {1\over 2} (p \g^{mn}\theta) + {1\over 2} (w \g^{mn}\lambda)
\ee
is the Lorentz current for all worldsheet variables except
for the non-minimal variables $(\lb_\alpha,\bar w^\alpha)$ and $(r_\alpha, s^\alpha)$. Since
the $b$-ghost is spacetime supersymmetric and a Lorentz scalar, $ \{ \oint dy\, q_\a(y),\, b(z) \} =0$ and 
 $ \left[ \oint dy\, (M^{mn}(y) + \bar M^{mn}(y)), \,b(z) \right] =0$
where
\be\label{nonminM}
\bar M^{mn} = {1\over 2} (\bar w\g^{mn} \l ) + {1\over 2} (s\g^{mn} r)
\ee
is the non-minimal contribution to the Lorentz current. So
\be\label{butoint}\bn
\pb b = \left[{1\over{2\pi i}} \oint dy\, V(y),\, b(z) \right] &=  \left[ {1\over{2\pi i}}\oint dy\, \left(q_\a(y) W^\a|_{\theta=0} -\frac12 M^{mn}(y)F_{mn}\right),\, b(z)\right]\\ &= \left[ {1\over{2\pi i}}\oint dy\, \frac12  \bar M^{mn}(y) F_{mn},\, b(z)\right].
\en\ee
But $\bar M_{mn} = -\frac12 \left\{Q, (s \g^{mn} \lb)\right\}$. So when $F_{mn}$ is constant, one learns from (\ref{butoint}) that $\pb b = \left[Q,\W_0\right]$ where
\be\label{Omegac}\bn
\W_0 &= -{1\over 4} F_{mn} \left[ {1\over{2\pi i}} \oint dy\,  \left(s(y)\g^{mn} \lb(y)\right), \,b(z)\right] = (\lb F)_\a \frac{\p b}{\p r_\a} \\
&= -\frac{\lb_\a\lb_\b}{(\bar\l\l)^2} {F_\g}^\a H^{[\b\g]}-2 \frac{\lb_\a\lb_\b r_\g}{(\lb\l)^3} {F_\d}^\a K^{[\b\g\d]} +3 \frac{\lb_\a\lb_\b r_\g r_\d}{(\lb\l)^4}  {F_\e}^\a L^{[\b\g\d\e]}.
\en\ee

\subsection{$\Omega$ in general super-Maxwell background}

Defining the singular OPE's of $b(y)$ with $V(z)$ as
\be\label{sing}
b(y) V(z) \to {{b_{1} V(z)}\over{(y-z)^3}}+{{b_{0} V(z)}\over{(y-z)^2}} +  {{b_{-1} V(z)}\over{y-z}}, 
\ee 
(\ref{comm}) implies that
\be\label{impc}
\pb b = - {1\over 2} \p^2 (b_1 V) + \p (b_0 V) - b_{-1} V
\ee
It will now be shown that the right-hand side of (\ref{impc}) is equal to $[Q,\W]$ where
\be\label{answer}
\W = b_{-1} b_0 V -{1\over 2} \p (b_{-1} b_1 V).
\ee

To compute $[Q,\W]$, use that $\{Q, b\} =T$ where 
\be\label{TV}
T(y) V(z) \to{{T_0 V(z)}\over{(y-z)^2} }+ {{T_{-1} V(z)}\over{y-z} }= {{V(z)}\over{(y-z)^2} }+ {{\p V(z)}\over{y-z} }
\ee
since $V$ is a primary field of conformal weight $+1$
when the Maxwell field $A_m$ appearing in (\ref {V}) is in Lorentz gauge, i.e. when $ {{\p}\over{\p x_m}}A_m (x) =0$. Furthermore, note that $[Q , V(z)]  = \p U(z)$ where 
$U(z) = \lambda^\alpha A_\alpha (x,\theta)$.
So 
\be\label{comput}
[Q,\W] = T_{-1} b_0 V - b_{-1} T_0 V + b_{-1} b_0 \p U -{1\over 2} \p (T_{-1} b_1 V - b_{-1} T_1 V
+ b_{-1} b_1 \p U)
\ee
\be
= \p (b_0 V) - b_{-1} V + \p (b_{-1} b_0 U) -{1\over 2} \p (\p(b_1 V) + b_{-1} \p (b_1 U) + 2 b_{-1} b_0 U)
\ee
where we used that $T_{-1}{\cal O} = \p{\cal O}$ and that
$\p(b_1 U) = b_1 \p U -2 b_0 U$. Cancelling the $\p(b_{-1} b_0 U)$ terms and using that
$b_1 U=0$, one obtains
\be
[Q,\W] = - {1\over 2} \p^2 (b_1 V) + \p (b_0 V) - b_{-1} V
\ee
which agrees with the right-hand side of (\ref{impc}).

Since the $b$-ghost of (\ref {b}) contains terms with poles of up to order $(\lb\l)^{-4}$, equation (\ref{answer}) implies that $\W$ can contain poles of up to order $(\lb\l)^{-8}$. However, note that $\W$ is only defined up to the equivalence $\W\sim\W + Q\Lambda$. And since
$\pb b = \left[{1\over{2\pi i}} \oint dy\, V(y), \,b(z) \right]$ only contains poles of up to order $(\lb\l)^{-4}$, it would
be surprising if an element in the cohomology of $\W$ cannot be chosen such that it only contains poles up to order  $(\lb\l)^{-4}$.
In the following subsection, a representative in the cohomology of $\W$ will be explicitly constructed which only contains poles up to order $(\lb\l)^{-4}$. 

\subsection{Explicit computation of $\pb b$}

Since $\pb b + \left[{1\over{2\pi i}} \oint dy\, \frac12 F_{mn} M^{mn}(y),\, b(z)\right]=0$ when $F_{mn}$ is constant,
it is convenient to define the covariant derivative 
\be\label{pb'}
\bar\nabla    \equiv \pb + {1\over{2\pi i}}  \oint dy\, \frac12 F_{mn} M^{mn}(y)
\ee
where $M^{mn}$ is
defined in (\ref{Mconst}).
The equations of motion of (\ref{eom1})-(\ref{eom2})  imply that
\be\label{rules}\bn
\bar\nabla    \l^\a      &= 0, \quad\bar\nabla    \w_\a      = 0, \\
\bar\nabla    d_\a       &= - \frac12 N_{mn} D_\a F^{mn}, \\
\bar\nabla    \p\th^\a &= - \P^m \p_m W^\a, \\
\bar\nabla   \P_m       &= d_\a \p_m W^\a + \frac12 N_{pq} \p_m F^{pq}.
\en\ee
With this notation, one may write for an arbitrary $F_{mn}$:
\be\label{db}\bn
\pb b &=  \left[{1\over{2\pi i}} \oint dy\, \frac12 F_{mn} \bar M^{mn}(y),\, b\right] + \bar\nabla    b = (\lb F)_\a \frac{\p b}{\p \lb_\a} + (r F)_\a \frac{\p b}{\p r_\a} + \bar\nabla    b\\
&= \left[Q, (\bar\l F)_\a \frac{\p b}{\p r_\a}\right] - \bar\l_\b \left[Q, {F^\b}_\a\right] \frac{\p b}{\p r_\a} + \bar\nabla    b,
\en\ee
where
\be\label{pb'b}
\bar\nabla    b = \frac{\bar\l_\a}{(\bar\l\l)}\, \bar\nabla    G^\a + \frac{\bar\l_\a r_\b}{(\bar\l\l)^2}\, \bar\nabla    H^{[\a\b]} - \frac{\bar\l_\a r_\b r_\g}{(\bar\l\l)^3}\, \bar\nabla    K^{[\a\b\g]} - \frac{\bar\l_\a r_\b r_\g r_\d}{(\bar\l\l)^4}\, \bar\nabla    L^{[\a\b\g\d]}.
\ee

We would now like to represent the right hand side of (\ref{db}) in the form $[Q,\W]$. In the next subsection \ref{fin}, we will construct such an $\W$. But before constructing $\W$, we will first find expressions for $\bar\nabla    G^\a, \bar\nabla    H^{[\a\b]}, \bar\nabla    K^{[\a\b\g]}, \bar\nabla    L^{[\a\b\g\d]}$ in terms of the spacetime supersymmetric operators $H^{\a\b}, K^{\a\b\g}, L^{\a\b\g\d}$ of
(\ref{nons}). The most direct method to find these expressions is to plug in the equations of motion of (\ref{rules}), and this method will be used in the appendix to find $\bar\nabla    G^\a$. However, a more efficient method which will be used here is to apply cohomology arguments based on the BRST structure of the equations of motion.


By acting with $Q$ on (\ref{db}), we find a relation
\be\label{beta}
\left\{Q, \bar\nabla    b - \bar\l_\b \l^\g D_\g {F^\b}_\a \frac{\p b}{\p r_\a} \right\} = 0.
\ee
This equation is polynomial in $r_\a$, and we can solve it order by order in $r_\a$. 

For terms which are zeroth order in $r_\a$, equation (\ref{beta}) gives
\be
\frac{\bar\l_\a}{(\bar\l\l)} \left\{ Q, \bar\nabla    G^\a \right\} = -\bar\l_\b \l^\g D_\g {F^\b}_\a \left( -\frac{\bar\l_\d \left[ Q, H^{[\a\d]}\right]}{(\bar\l\l)^2} \right) = \frac{\bar\l_\b}{(\bar\l\l)} \l^\g D_\g {F_\a}^\b G^\a,
\ee
which implies that 
\be\label{a}\bn
\left\{ Q, \bar\nabla   G^\a \right\} = \l^\b D_\b {F_\g}^\a G^\g &= D_\b {F_\g}^\a \left[Q, H^{\b\g}\right] \\&= \left\{ Q, -D_\b {F_\g}^\a H^{\b\g} \right\} + \l^\b D_\b D_\g {F_\d}^\a H^{\g\d}.
\en\ee
We have used the first of the relations (\ref{cep}) and that
\begin{gather}
g^{((\a\b))} D_\a {F_\b}^\g = g^{((\a\b))} D_\a D_\b W^\g = 0 
\end{gather}
since $g^{((\a\b))}$ is symmetric and gamma matrix traceless.
We can rewrite the last term in (\ref{a}) using the relation for $\{Q,K^{\a\b\g}\}$ in (\ref{cep}). Since $h_1^{((\a\b))\g}$ and $h_2^{\a((\b\g))}$ are symmetric and gamma matrix traceless in the pairs of indices shown, both of them vanish when contracted with $D_\a D_\b {F_\g}^\d$, and we are left with
\be
\left\{ Q, \bar\nabla   G^\a + D_\b {F_\g}^\a H^{\b\g} \right\} = D_\b D_\g {F_\d}^\a \left\{ Q, K^{\b\g\d} \right\}.
\ee
We repeat the same procedure once again, transforming the right hand side:
\be
D_\b D_\g {F_\d}^\a \left\{ Q, K^{\b\g\d} \right\} = \left\{ Q, D_\b D_\g {F_\d}^\a K^{\b\g\d} \right\} - \l^\b D_\b D_\g D_\d {F_\e}^\a K^{\g\d\e}.
\ee
The last term may be rewritten with the help of the next relation in (\ref{cep}). We get:
\be\bn
- \l^\b D_\b D_\g D_\d {F_\e}^\a K^{\g\d\e} &= -D_\b D_\g D_\d {F_\e}^\a \left[ Q, L^{\b\g\d\e} \right] =\\
&= \left\{ Q, D_\b D_\g D_\d {F_\e}^\a L^{\b\g\d\e} \right\} - \l^\b D_\b D_\g D_\d D_\e {F_\z}^\a L^{\g\d\e\z}.
\en\ee
The last term here vanishes since $L^{\a\b\g\d}$ satisfies the last equation in (\ref{cep}).

To summarize, we have shown that 
\be\label{otvet-G}
\left\{ Q, \bar\nabla   G^\a + D_\b {F_\g}^\a H^{\b\g} - D_\b D_\g {F_\d}^\a K^{\b\g\d} - D_\b D_\g D_\d {F_\e}^\a L^{\b\g\d\e} \right\}=0.
\ee
Since $\bar\nabla   G^\a + \ldots$ is an operator of conformal weight $2$, it must be in the trivial cohomology class of $Q$, i.e. there must exist some $\F^\a$ of ghost-number $-1$ such that
\be
\bar\nabla   G^\a + D_\b {F_\g}^\a H^{\b\g} - D_\b D_\g {F_\d}^\a K^{\b\g\d} - D_\b D_\g D_\d {F_\e}^\a L^{\b\g\d\e} = \left[ Q, \F^\a \right].
\ee
But since there are no ghost number $-1$ operators which are invariant under the gauge symmetry of (\ref{gaugew}), $\F^\a$ must be zero. 

Essentially the same steps can be repeated for the terms of higher order in $r_\a$ in the main equation (\ref{beta}). The corresponding equations, after algebraic simplifications, are given by:
\be\label{poryadki}\bn
\lb_\a r_\b &\left(\left[ Q, \bar\nabla    H^{[\a\b]} \right] - \l^{\left[\a\right.} \bar\nabla    G^{\left.\b\right]} + \l^\g D_\g {F_\d}^{\left[\a\right.} H^{\left.\b\d\right]}\right) = 0, \\
\lb_\a r_\b r_\g &\left(\left\{ Q, \bar\nabla    K^{[\a\b\g]} \right\} - \l^{\left[\a\right.} \bar\nabla    H^{\left.\b\g\right]} - \l^\d D_\d {F_\e}^{\left[\a\right.} K^{\left.\b\g\e\right]}\right) = 0, \\
\lb_\a r_\b r_\g r_\d &\left(\left[ Q, \bar\nabla    L^{[\a\b\g\d]} \right] - \l^{\left[\a\right.} \bar\nabla    K^{\left.\b\g\d\right]} + \l^\e D_\e {F_\z}^{\left[\a\right.} L^{\left.\b\g\d\z\right]}\right) = 0, \\
\lb_\a r_\b r_\g r_\d r_\e &\,\l^{\left[\a\right.} \bar\nabla    L^{\left.\b\g\d\e\right]}  = 0.
\en\ee
Note that the last equation is satisfied identically using the fact that $\bar\nabla   \l^\a =0$ and the antisymmetrized version of the last equation in (\ref{cep}). One also has to use the latter relation in order to transform the $r^3$ equation to the form given in the third line of (\ref{poryadki}).
Solving these equations follows the same scheme as explained above for (\ref{a}). 

The result is 
\begin{gather}
\label{d'G}
\bar\nabla   G^\a = -D_\b {F_\g}^\a H^{\b\g} + D_\b D_\g {F_\d}^\a K^{\b\g\d} + D_\b D_\g D_\d {F_\e}^\a L^{\b\g\d\e},\\
\label{d'H}
\bar\nabla   H^{[\a\b]} = -D_\g {F_\d}^{\left[\a\right.} \tilde K^{\left.\b\right]\g\d} - D_\g D_\d {F_\e}^{\left[\a\right.} \tilde L^{\left.\b\right]\g\d\e}, \\
\label{d'K}
\bar\nabla   K^{[\a\b\g]} = -\frac12 D_\d {F_\e}^{\left[\a\right.} \tilde{\tilde L}^{\left.\b\g\right]\d\e},\quad \bar\nabla   L^{[\a\b\g\d]} = 0,
\end{gather}
where
\begin{gather}
\label{deftildes}
\tilde K^{\a\b\g} = K^{\a\b\g} - K^{\b\a\g} + K^{\b\g\a},\\
\tilde L^{\a\b\g\d} = L^{\a\b\g\d} - L^{\b\a\g\d} + L^{\b\g\a\d} - L^{\b\g\d\a},\\
\tilde{\tilde L}^{\a\b\g\d} = \tilde L^{\a\b\g\d} - \tilde L^{\a\g\b\d} + \tilde L^{\a\g\d\b}.
\end{gather}
It will be shown in the appendix that the last term in (\ref {d'G}), $D_\b D_\g D_\d {F_\e}^\a L^{\b\g\d\e}$, is identically zero. However, it is convenient for intermediate steps to include this term. 

\subsection{Explicit construction of $\W$}
\label{fin}

Let us examine how the function $\W_0$ given in (\ref{Omegac}) should be extended in order to incorporate nonzero derivatives of $F_{mn}$. Consider the following extension of the first term in (\ref{Omegac}) by terms that depend on $DF$ and $D^2 F$:
\be\label{W1}
\W_1 = \frac{\lb_\a\lb_\b}{(\bar\l\l)^2} \left( -{F_\g}^\a H^{[\b\g]} + D_\g {F_\d}^\a \tilde K^{\b\g\d} + D_\g D_\d {F_\e}^\a \tilde L^{\b\g\d\e} \right).
\ee
It will now be shown that, to zeroth order in $r$, the BRST image of this function is
\be\label{pbb}
\left.\left[ Q, \W_1 \right]\right|_{r^0}= \frac{\bar\l F G}{(\bar\l\l)} - \frac{\left(\bar\l F\l\right)\left(\bar\l G\right)}{(\bar\l\l)^2} + \frac{\bar\l_\a}{\bar\l\l} \,\bar\nabla    G^\a = \left. \pb b \right|_{r^0}
\ee
where we have used  (\ref{db}).
To check this, first calculate the BRST variations of $\tilde K, \tilde L$ using (\ref{cep}):
\be
\left\{ Q, \tilde K^{\a\b\g} \right\} = \l^\a H^{\b\g} - \l^\b H^{[\a\g]} + h_1^{((\b\g))\a} + h_2^{\a((\b\g))}, \\
\ee
\be\bn
\left[ Q, \tilde L^{\a\b\g\d} \right] = \l^\a K^{\b\g\d} - \l^\b \tilde K^{\a\g\d} &+ k_1^{((\b\g))\a\d} - k_1^{((\b\g))\d\a} + k_2^{\a((\b\g))\d} \\&- k_2^{\b((\g\d))\a} + k_3^{\a\b((\g\d))} - k_3^{\b\a((\g\d))}.
\en\ee
Note that the contractions of $\tilde K, \tilde L$ with $DF$, $D^2 F$ within (\ref{W1}) are such that all $h_i$, $k_j$ terms that will appear in the BRST variation of $\W_1$ cancel. Therefore we have for the BRST variations of different terms inside the bracket in (\ref{W1})
\be\bn
\left[ Q, -{F_\g}^\a H^{[\b\g]} \right] &= \l^\d D_\d {F_\g}^\a H^{[\g\b]} + {F_\g}^\a \left( \l^\g G^\b - \l^\b G^\g \right),\\
\left[ Q, D_\g {F_\d}^\a \tilde K^{\b\g\d} \right] &= \l^\e D_\e D_\g {F_\d}^\a \tilde K^{\b\g\d} - D_\g {F_\d}^\a \left( \l^\b H^{\g\d} - \l^\g H^{[\b\d]} \right) ,\\
\left[ Q, D_\g D_\d {F_\e}^\a \tilde L^{\b\g\d\e} \right] &= \l^\z D_\z D_\g D_\d {F_\e}^\a \tilde L^{\b\g\d\e} \\&+ D_\g D_\d {F_\e}^\a \left( \l^\b K^{\g\d\e} - \l^\g \tilde K^{\b\d\e}\right).
\en\ee
Putting these results together and multiplying by $\frac{\lb_\a\lb_\b}{(\lb\l)^2}$, we recover (\ref{pbb}), with $\bar\nabla   G^\a$ given by (\ref{d'G}).

One can similarly check that the modification of the remaining terms in (\ref{Omegac}) to incorporate nonvanishing $DF$ is the following:
\begin{gather}
\label{W2}
\W_2 = \frac{\lb_\a\lb_\b r_\g}{(\lb\l)^3} \left( -2 {F_\d}^\a K^{[\b\g\d]} + 2 D_\d {F_\e}^\a \tilde{\tilde L}^{\g\b\d\e} \right),\quad \W_3 = \frac{\lb_\a\lb_\b r_\g r_\d}{(\lb\l)^4} 3 {F_\e}^\a L^{[\b\g\d\e]}
\end{gather}
($\W_3$ is not modified). Together with (\ref{W1}), these form a complete inverse BRST image of $\pb b$, at all orders in $r_\a$ and with corrections due to the derivatives of the background field strength:
\be
\pb b = \left[{1\over{2\pi i}} \oint dy\, \frac12 F_{mn}  \bar M^{mn}(y),\, b\right] + \bar\nabla    b  = \left[ Q, \W_1 + \W_2 + \W_3 \right] 
\ee
where $\bar\nabla   b$ is given by (\ref{pb'b}) and (\ref{d'G})-(\ref{d'K}).

So using the expressions of (\ref {W1}) and (\ref{W2}), one finds $\pb b = [Q,\W]$ where
\be\label{final}
\W = \W_0 +  \frac{\lb_\a\lb_\b}{(\bar\l\l)^2} \left(D_\g {F_\d}^\a \tilde K^{\b\g\d} + D_\g D_\d {F_\e}^\a \tilde L^{\b\g\d\e} \right)  +2
\frac{\lb_\a\lb_\b r_\g}{(\lb\l)^3} D_\d {F_\e}^\a \tilde{\tilde L}^{\g\b\d\e} 
\ee
and $\W_0$ is defined in (\ref{Omegac}) and $\tilde K^{\b\g\d}$, $\tilde L^{\b\g\d\e}$ and $\tilde{\tilde L}^{\g\b\d\e}$ are defined in
(\ref{deftildes}).

\section*{Acknowledgements}
\label{ack}
We would like to thank Ido Adam, Oscar Chacaltana, Sebastian Guttenberg, Renann Jusinskas, and Andrei Mikhailov for useful discussions. NB would also like
to thank CNPq grant 300256/94-9
and FAPESP grants 09/50639-2 and 11/11973-4 for partial financial support, and IB would like to thank FAPESP grant 2011/00157-1 for financial support.

\appendix



\section{Computation of $\bar\nabla    G^\a$}
\label{app1}

Here we show by explicit computation that the expression which is acted upon by $Q$ in (\ref{otvet-G}) is in fact identically zero:
\be\label{c1}
\bar\nabla   G^\a + D_\b {F_\g}^\a H^{\b\g} - D_\b D_\g {F_\d}^\a K^{\b\g\d} - D_\b D_\g D_\d {F_\e}^\a L^{\b\g\d\e} = 0.
\ee
Using the action of $\bar\nabla   $ on elementary fields defined in (\ref{rules}), one finds:
\be\label{c2}\bn
\bar\nabla    G^\a &= \frac12 \left( d_\b \p_m W^\b + \frac12 N_{pq} \p_m F^{pq} \right) (\g^m d)^\a -\frac14 \P^m \g_m^{\a\b} N_{pq} D_\b F^{pq} \\&+ \frac14 N_{mn} {(\g^{mn})^\a}_\b \P^k \p_k W^\b + \frac14 J \P^k \p_k W^\a.
\en\ee
The next term in (\ref{c1}) can be calculated using $D_\b {F_\g}^\a = \frac12 {(\g^{pq})_\g}^\a \p_p (\g_q W)_\b$ and the expression for $H^{\b\g}$ \cite{Berkovits:2004px}:
\be
H^{\b\g} = \frac{1}{16} N^{mn} \P^k (\g_m \h_{nk} + \g_{mnk})^{\b\g} - \frac{1}{32} \g_m^{\b\g} J \P^m -\frac{1}{16} (\g_m d)^\b (\g^m d)^\g.
\ee
Multiplying the two expressions and using gamma matrix identities in order to simplify the result, one finds that $D_\b {F_\g}^\a H^{\b\g}$ cancels most of the terms in (\ref{c2}):
\be\bn
D_\b {F_\g}^\a H^{\b\g} = &-\frac12 ( d \p_m W ) (\g^m d)^\a +\frac14 \P^m \g_m^{\a\b} N_{pq} D_\b F^{pq} \\&- \frac14 N_{mn} {(\g^{mn})^\a}_\b \P^k \p_k W^\b - \frac14 J \P^m \p_m W^\a.
\en\ee
To calculate $D_\b D_\g {F_\d}^\a K^{\b\g\d}$ we use the following expressions:
\be
D_\b D_\g {F_\d}^\a = \frac18 {(\g^{pq})_\d}^\a (\g_{mn} \g_q)_{\b\g} \p_p F^{mn},
\ee
\be
K^{\b\g\d} = \frac{1}{128} k_1^{\b\g\d} + \frac{1}{192} k_2^{\b\g\d}  - \frac{1}{48} k_3^{\b\g\d} - \frac{1}{192} k_4^{\b\g\d} - \frac{1}{192} k_5^{\b\g\d},
\ee
where
\be\bn
k_1^{\b\g\d} &= \g_m^{\g\d} (\g^m d)^\b J, \\
k_2^{\b\g\d} &= \g_m^{\g\d} (\g_n d)^\b N^{mn},\\
k_3^{\b\g\d} &= \g_m^{\b\g} (\g_n d)^\d N^{mn}, \\
k_4^{\b\g\d} &= \g_{mnk}^{\b\g} (\g^m d)^\d N^{nk},\\
k_5^{\b\g\d} &= \g_{mnk}^{\g\d} (\g^m d)^\b N^{nk}.
\en\ee
The relation for $D_\b D_\g {F_\d}^\a$ can be easily derived using (\ref{max}). The explicit form of $K^{\b\g\d}$ is derived in \cite{Oda:2007ak}. Contracting the spinorial indices and using various gamma matrix identities to simplify the expressions, we get the following:
\be\bn
D_\b D_\g {F_\d}^\a \,k_1^{\b\g\d} &= 0, \\
D_\b D_\g {F_\d}^\a \,k_2^{\b\g\d} &= -2 N^{km} \p_m F^{pq} (d\g_{kpq})^\a - 2 N_{pq} \p_m F^{pq} (\g^m d)^\a, \\
D_\b D_\g {F_\d}^\a \,k_3^{\b\g\d} &= -2 N^{km} \p_m F^{pq} (d\g_{kpq})^\a - 2 N_{pq} \p_m F^{pq} (\g^m d)^\a, \\
D_\b D_\g {F_\d}^\a \,k_4^{\b\g\d} &= -4 N^{km} \p_m F^{pq} (d\g_{kpq})^\a - 28 N_{pq} \p_m F^{pq} (\g^m d)^\a, \\
D_\b D_\g {F_\d}^\a \,k_5^{\b\g\d} &= 10 N^{km} \p_m F^{pq} (d\g_{kpq})^\a - 14 N_{pq} \p_m F^{pq} (\g^m d)^\a. \\
\en\ee
Adding up the contributions, we get
\be
D_\b D_\g {F_\d}^\a K^{\b\g\d} = \frac14 N_{pq} \p_m F^{pq} (\g^m d)^\a.
\ee
This cancels precisely with the remaining term in (\ref{c2}).

To prove (\ref{c1}) it remains to show that the last term $D_\b D_\g D_\d {F_\e}^\a L^{\b\g\d\e}$ vanishes. Using a generic expression for $L^{\b\g\d\e}$ \cite{Berkovits:2004px} with undetermined coefficients,
\be
L^{\b\g\d\e} = (l_1)^{\b\g\d\e} JJ + (l_2)^{\b\g\d\e}_{mn} JN^{mn} + (l_3)^{\b\g\d\e}_{mnkl} N^{mn} N^{kl},
\ee
one can show that every term vanishes independently due to the symmetry properties or using the equations of motion. In order to see this write $D^3 F$ as follows,
\be
D_\b D_\g D_\d {F_\e}^\a = D_\b D_\g D_\d D_\e W^\a = (X_{\b\g\d\e}^{mn})^\a_\z \,\p_m \p_n W^\z.
\ee
where $(X_{\b\g\d\e}^{mn})^\a_\z$ is a Lorentz-invariant tensor.
This should be possible since the only physical fields are the gluon and gluino, and any onshell gauge-invariant superfield can be expressed in terms of them and their spacetime derivatives. So we have
\be\bn
D_\b D_\g D_\d {F_\e}^\a L^{\b\g\d\e} = \Bigl[ JJ\, (X\cdot l_1)^{\a\,mn}_\b &+ J N_{kl}(X\cdot l_2)^{\a\,mnkl}_\b  \\  &+ N_{kl} N_{pq}(X\cdot l_3)^{\a\,mnklpq}_\b\Bigr]\p_m\p_n W^\b.
\en\ee
Using Fierz identities, expand $(X\cdot l_i)^{\a\,m\ldots}_\b$ in terms of $\d^\a_\b, {(\g^{mn})^\a}_\b$, and ${(\g^{mnkl})^\a}_\b$:
\be\bn
(X\cdot l_1)^{\a\,mn}_\b &= c_1 \h^{mn} \d^\a_\b + c_2 {(\g^{mn})^\a}_\b,\\
(X\cdot l_2)^{\a\,mnkl}_\b &= c_3 \h^{mn} \h^{kl} \d^\a_\b + c_4 {(\g^{mn})^\a}_\b \h^{kl} \\&+ c_5 {(\g^{mnkl})^\a}_\b + \mathrm{permutations},\\
(X\cdot l_3)^{\a\,mnklpq}_\b &= c_6 \h^{mn} \h^{kl} \h^{pq} \d^\a_\b + c_7 {(\g^{mn})^\a}_\b \h^{kl} \h^{pq} \\&+ c_5 {(\g^{mnkl})^\a}_\b \h^{pq} + \mathrm{permutations},
\en\ee
with some constant coefficients $c_i$. Using (anti)symmetry of $N_{kl}$ and $\p_m \p_n W^\a$ allows to remove many of the terms, and to write most of the others in a unified manner. The only surviving terms that are not the same are:
\be\bn
JJ\, (X\cdot l_1)^{\a\,mn}_\b \,\p_m\p_n W^\b \rightarrow & \,JJ\,\h^{mn}\, \d^\a_\b\, \p_m \p_n W^\b,\\
J N_{kl}\,(X\cdot l_2)^{\a\,mnkl}_\b \,\p_m\p_n W^\b \rightarrow & \,J N_{kl}\, \h^{mk} \h^{nl} \d^\a_\b\, \p_m \p_n W^\b,\\ &\,J N_{kl}\,{(\g^{kl})^\a}_\b\, \h^{mn} \p_m \p_n W^\b,\\ & \,J N_{kl}\,{(\g^{mk})^\a}_\b\, \h^{nl} \p_m \p_n W^\b,\\
N_{kl} N_{pq}\,(X\cdot l_3)^{\a\,mnklpq}_\b \,\p_m\p_n W^\b \rightarrow &\, N_{kl} N_{pq}\, {(\g^{mklp})^\a}_\b\, \h^{nq} \,\p_m\p_n W^\b,\\ &\, N_{kl} N_{pq}\, {(\g^{mk})^\a}_\b\, \h^{lp}\h^{nq} \,\p_m\p_n W^\b,\\ &\, N_{kl} N_{pq}\, {(\g^{kl})^\a}_\b\, \h^{mq}\h^{np} \,\p_m\p_n W^\b,\\ &\, N_{kl} N_{pq}\, {(\g^{kp})^\a}_\b\, \h^{ml}\h^{nq} \,\p_m\p_n W^\b,\\ &\, N_{kl} N_{pq}\,  \h^{mk}\h^{np}\h^{lq} \,\p_m\p_n W^\a.
\en\ee
It is easy to see that all of these terms vanish either due to symmetry properties, or using field equations $\g^m_{\a\b}\p_m W^\b = 0$, $\p^m \p_m W^\a = 0$.

\pagebreak

\bibliographystyle{utphys}
\bibliography{bib}

\end{document}